\documentclass{jpsj-suppl}
\usepackage{txfonts} 
\usepackage{graphicx}
\usepackage{bm}
\usepackage{cite}

\title{Neutron Scattering Study on Commensurate and Incommensurate Antiferromagnetic Phases in UPd$_{\bm 2}$Si$_{\bm 2}$ under Uniaxial Stress}

\author{Takahiro \textsc{Nakada}$^{1,\dag}$, Makoto \textsc{Yokoyama}$^{1,\ddag}$, Chihiro \textsc{Tabata}$^{2}$, Hideki \textsc{Igarashi}$^{2}$, \\Hiroyuki \textsc{Hidaka}$^{2}$, Hiroshi \textsc{Amitsuka}$^{2}$, Kenichi \textsc{Tenya}$^{3}$ and Taku J. \textsc{Sato}$^{4}$}

\inst{$^{1}$Faculty of Science, Ibaraki University, Mito 310-8512 \\
$^{2}$Graduate School of Science, Hokkaido University, Sapporo 060-0810 \\
$^{3}$Faculty of Education, Shinshu University, Nagano 380-8544 \\
$^{4}$Neutron Science Laboratory, Institute for Solid State Physics, The University of Tokyo, Tokai 319-1106 \\}

\email{$^\dag$10nm163n@mcs.ibaraki.ac.jp, $^\ddag$makotti@mx.ibaraki.ac.jp}

\recdate{\today}

\abst{The nature of competition between incommensurate (IC) and commensurate (C) antiferromagnetic (AF) orders in UPd$_2$Si$_2$ was investigated by performing elastic neutron scattering experiments under uniaxial stress $\sigma$. It is found that applying $\sigma$ along tetragonal [010] direction reduces the IC-AF order, and then stabilizes the C-AF order. The transition temperature from IC- to C-AF phases $T_{\rm Nl}$ is enhanced from 109 K ($\sigma=0$) to 112.5 K (0.8 GPa), while the onset of IC-AF transition $T_{\rm Nh}$ is unchanged from 132 K under $\sigma$. In addition, $c$-axis component $q_z$ of the IC-AF modulation at 115 K also increases from 0.736 ($\sigma=0$) to 0.747 (0.8 GPa). The magnitude of C-AF moment at 5 K is estimated to be 2.2 $\mu_{\rm B}/{\rm U}$ in the entire $\sigma$ range presently investigated ($\sigma \le 0.8\ {\rm GPa}$). These features are similar to those obtained from the investigations using hydrostatic pressure $p$, indicating that applications of $p$ and $\sigma\,||\,[010]$ commonly induce the crystal strains which inherently affect a delicate balance of frustrated magnetic interactions between uranium 5f moments.}

\kword{UPd$_2$Si$_2$, antiferromagnetism, uniaxial stress, neutron scattering}

\begin{document}
\maketitle

\section{Introduction}
The ternary uranium compound UPd$_2$Si$_2$ (the ThCr$_2$Si$_2$-type body centered tetragonal structure) displays a variety of antiferromagnetic (AF) modulations of uranium 5f moments in the ordered states \cite{rf:Palstra86,rf:Shemirani93,rf:Collins93,rf:Honma98,rf:Plackowski11,rf:Wermeille}. In UPd$_2$Si$_2$, an incommensurate (IC) AF order composed of a longitudinal sine-wave with a propagation vector of $q_1= (0,0,0.73)$ develops below $T_{\rm Nh} = 132-138\ {\rm K}$, and it is replaced by a commensurate (C) AF order with $q_2 = (0,0,1)$ below $T_{\rm Nl} = 108\ {\rm K}$ via a first-order phase transition. Furthermore, applying magnetic field along the tetragonal $c$ axis suppresses both the AF phases, and then generates the other anti-phase AF structure with a modulation of $q_3 = (0,0,2/3)$ above 6 T. These features on the AF modulations are considered to be ascribed to a frustration of different inter-site interactions between the 5f moments with nearly localized characteristics \cite{rf:Honma98}. In fact, these features are suggested to be fairly well reproduced by the theoretical calculation based on the axial-next-nearest-neighbor Ising (ANNNI) model \cite{rf:Honma98}. In such a situation, it is expected that the competition among the magnetic interactions is sensitively affected by the crystal strains that give rise to the variations of the distances between uranium ions. To investigate the relationship between the AF modulations and the crystal strains, we have performed elastic neutron scattering experiments under uniaxial stress for UPd$_2$Si$_2$.

\section{Experiment Details}
A single crystal of UPd$_2$Si$_2$ was grown by a Czochralski pulling method using a tri-arc furnace, and a plate-shaped sample with a base of the (010) plane (dimensions: 13.1mm$^2$$ \times$ 1mm) was cut out from the ingot by means of a spark erosion. The sample was then mounted between pistons (Be-Cu alloy) in a constant-load uniaxial-stress apparatus \cite{rf:Kawarazaki02}, and cooled down to 1.5 K in a pumped $^4$He cryostat. The uniaxial stress was applied along the [010] directions (equivalent to the [100] direction in the tetragonal symmetry) up to 0.8 GPa. The elastic neutron scattering experiments were performed in the triple-axis spectrometer ISSP-GPTAS located at the JRR-3 reactor of JAEA, Tokai. We chose the neutron momentum of $k=2.67~$\AA$^{-1}$, and used a combination of 40'-40'-40'-80' collimators and two pyrolytic graphite filters. The scans were performed in the $(h0l)$ scattering plane. Applying $\sigma$ along the [010] direction up to 0.8 GPa is expected to enhance the $a$- and $c$-axes lattice parameters by a ratio of $\sim 10^{-4}$ \cite{rf:Yokoyama05}. In the present investigation, however, such variations cannot be detected due to limitation of instrumental resolution ($\sim 10^{-2}-10^{-3}$).
\begin{figure}[tbp]
\begin{center}
\includegraphics[keepaspectratio,width=0.75\textwidth]{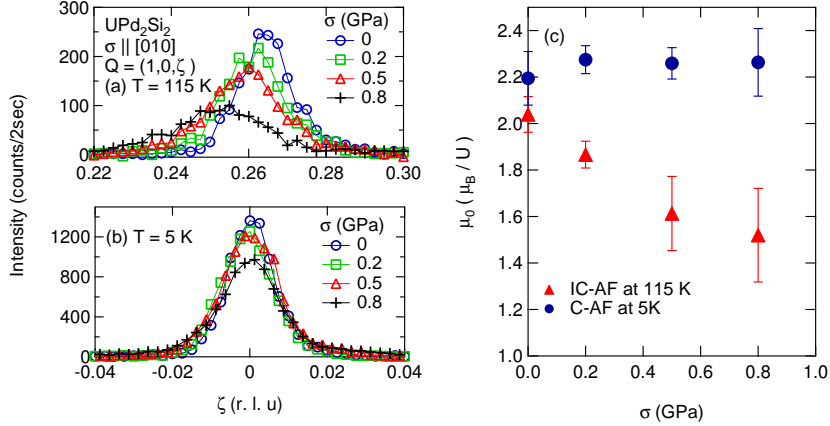}
\end{center}
\vspace{-10pt}
  \caption{
Uniaxial-stress variations of the AF Bragg-peak profiles obtained from the $(1,0,\zeta)$ line scans (a) around $\zeta=0.264$ at 115 K, and (b) around $\zeta=0$ at 5 K for UPd$_2$Si$_2$. The volume-averaged AF moments for the C-AF and IC-AF orders are shown in (c). }
\end{figure}
\section{Results and Discussion}
 Figure 1(a) and (b) show $\sigma$ variations of IC- and C-AF Bragg-peak profiles for the momentum transfer $Q=(1,0,\sim 0.264) \ (\equiv Q_1)$ at 115 K and $Q=(1,0,0) \ (\equiv Q_2)$ at 5 K, respectively.  At $\sigma=0$, clear Bragg peaks originating from the IC-AF order with the propagation vector $q_1=(0,0,0.736(2))$ are observed at 115 K. As temperature is lowered, the IC-AF Bragg peaks disappear and then the C-AF Bragg peaks develop. By applying $\sigma$, it is found that both the intensity and the IC component $\delta$ of the peak position $(1,0,\delta)$ for the IC-AF Bragg peak at $Q_1$ are significantly reduced, although the C-AF Bragg peaks at $Q_2$ are roughly insensitive to $\sigma$. We also observed that the widths of the IC-AF Bragg peak for $\sigma=0.8\ {\rm GPa}$ are somewhat larger than the instrumental resolution estimated from the widths of nuclear Bragg peaks, while those for $\sigma \le 0.5\ {\rm GPa}$ are resolution-limited. This broadening occurs in the entire  temperature range of the IC-AF phase appearing, and is nearly independent of temperature. In addition, such a broadening occurs only in the IC-AF Bragg peaks whose positions markedly vary by applying $\sigma$, and is not seen in the other nuclear and C-AF Bragg reflections presently measured. We thus consider that the broadening may be caused by a distribution of the IC wave vectors in the sample generated by inhomogeneity of $\sigma$ in high-$\sigma$ region (roughly $\pm 0.2\ {\rm GPa}$ at $0.8\ {\rm GPa}$), not by short-ranged metastable AF clusters presumably formed on the verge of the first-order phase boundary at $T_{\rm Nl}$. 
\begin{figure}[tbp]
\begin{center}
\includegraphics[keepaspectratio,width=0.75\textwidth]{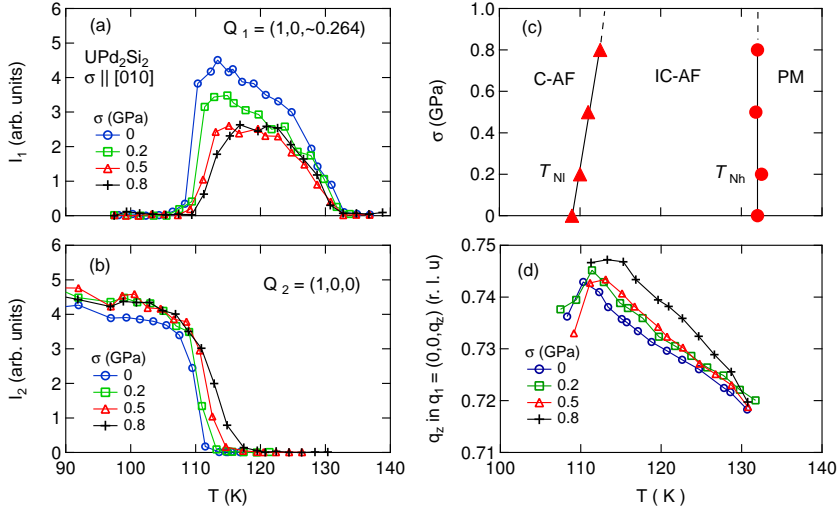}
\end{center}
\vspace{-10pt}
  \caption{
Temperature variations of the AF Bragg-peak intensities for the (a) IC- and (b) C-AF orders, and (d) the $c$ axis component of the ordering vector in the IC-AF phase of UPd$_2$Si$_2$. In (c), the AF transition temperatures $T_{\rm Nh}$ and $T_{\rm Nl}$ under $\sigma$ for IC- and C-AF orders are plotted. 
}
\end{figure}

In Fig.\ 1(c), $\sigma$ variations of volume-averaged AF moments $\mu_{\rm 0}$ for the IC-AF order (115 K) and the C-AF order (5K) are plotted. The $\mu_{\rm 0}$ values are estimated from the Bragg-peak intensities at $Q_1$ and $Q_2$ normalized by the intensities of the nuclear (101) Bragg reflections as a reference. The $|Q|$ dependence of the magnetic scattering amplitude is assumed to be proportional to the magnetic form factor of the U$^{4+}$ ion \cite{rf:Freeman76}. The $\mu_0$ values for the C-AF order under $\sigma$ are unchanged from $2.2(1)\ \mu_{\rm B}/{\rm U}$ at ambient pressure, but those for the IC-AF order are linearly reduced from $2.0(1)\ \mu_{\rm B}/{\rm U}$ ($\sigma=0$) to $1.5(2)\ \mu_{\rm B}/{\rm U}$ ($0.8\ {\rm GPa}$).

Displayed in Fig.\ 2(a) and (b) are the IC- and C-AF Bragg-peak intensities $I_1$ and $I_2$ estimated at $Q_1$ and $Q_2$, respectively, plotted as a function of temperature. For $\sigma=0$, $I_1$ starts increasing at 132 K as temperature is decreased, and shows a tendency to saturate below $\sim 115\ {\rm K}$. The phase transition from IC- to C-AF orders is clearly observed at $\sim 109\ {\rm K}$ in both $I_1$ and $I_2$, where $I_1$ discontinuously drops to zero and $I_2$ increases sharply. Similar variations of $I_1$ and $I_2$ are also observed under $\sigma$. However, the discontinuous changes of $I_1$ and $I_2$ ascribed to the IC- to C-AF transition occur in wider temperature range by applying $\sigma$, probably due to the inhomogeneity of strain in the sample. We here define the transition temperature $T_{\rm Nh}$ from the onset of $I_1$, and $T_{\rm Nl}$ as a midpoint of the step-like development in $I_2$. Although $T_{\rm Nh}$ remains nearly unaffected from $\sigma$, $T_{\rm Nl}$ linearly increases with increasing $\sigma$ at a rate of $\partial T_{\rm Nl}/\partial \sigma =4.3\ {\rm K/GPa}$ [Fig.\ 2(c)]. Simple extrapolations of the $\sigma$-variations of $T_{\rm Nh}$ and $T_{\rm Nl}$ yield that they meet with each other at $5.4\ {\rm GPa}$.

In Fig.\ 2(d), we show temperature dependence of the $c$-axis component $q_z(T)$ in $q_1$ obtained from the Bragg reflections at $Q_1$. The overall features of $q_z(T)$ do not change by the application of $\sigma$: $q_z$ linearly increases with decreasing temperature, and shows a maximum at $T^*$ ($\sim 110\ {\rm K}$ at $\sigma=0$), followed by a reduction with further decreasing temperature. However, both $T^*$ and the magnitude of $q_z$ above $T^*$ are found to be slightly enhanced by the compression. At $\sigma=0.8\ {\rm GPa}$, $q_z$ reaches 0.747(3) at 115 K, and the determination of $T^*$ becomes difficult because the cusp in the $q_z(T)$ curve are obscured.

 We have found that applying $\sigma$ along the [010] direction stabilizes the C-AF phase through tuning the competition between the C- and IC-AF orders. This tendency is clearly indicated by the increases of $T_{\rm Nl}$ and $q_z$ under $\sigma$. Interestingly, the presently obtained $\sigma-T$ phase diagram closely resembles the pressure $p$ versus temperature phase diagram proposed by the resistivity measurements \cite{rf:Quirion98,rf:Hidaka11}. In general, it is expected that compressions using $p$ and $\sigma\,||\,[010]$ yield the different types of strain. In particular, the $c$-axis lattice parameter should be reduced by applying $p$, while it may be slightly enhanced by $\sigma$. This suggests that the strain other than the variation of $c$-axis lattice parameter, such as the $a$-axis lattice parameter or tetragonal $c/a$ ratio, plays an crucial role in the frustration among the different inter-site interactions of uranium 5f moments. We also wish to stress that $\sigma$ along the [010] direction induces the tetragonal symmetry-breaking strain of the $x^2-y^2$ type. The relationship between the symmetry-breaking strains and the AF orders is not clear in the present stage. We thus need to investigate effects of $\sigma$ along the other directions on the AF orders. On the other hand, both the $\mu_0$ at 5 K and $T_{\rm Nh}$ values are not influenced by $\sigma$, suggesting that the characteristics of uranium 5f electrons, such as a valence and an AF condensation energy, are basically unchanged under $\sigma$. 

\section{Summary}
Our elastic neutron scattering experiments under $\sigma\  (||\, [010])$ for UPd$_2$Si$_2$ revealed that the compression with $\sigma$ favors the evolution of the C-AF order rather than the IC-AF order. The transition temperature $T_{\rm Nl}$ from the IC- to C-AF states and the $c$-axis IC modulation $q_z$ of the IC-AF order increase by applying $\sigma$, while the staggered moment of the C-AF order and the IC-AF transition temperature $T_{\rm Nh}$ are insensitive to $\sigma$. We suggest from the resemblance between the $\sigma-T$ and $p-T$ phase diagrams that there are couplings between the AF orders and crystal strains commonly induced by the different types of compression, and they govern the frustration of the AF states. To clarify a detail of such couplings, we plan to perform the neutron scattering experiments by applying $\sigma$ along the other directions.

\section*{Acknowledgment}
We thank T. Asami, R. Sugiura and Y. Kawamura for technical support on neutron scattering measurements. This work was partly supported by a  Grant-in-Aid for Scientific Research on Innovative Areas ``Heavy Electrons" (No.23102703) from the Ministry of Education, Culture, Sports, Science and Technology of Japan.

\end{document}